\documentclass[aps,prb,manuscript]{revtex4-1}
\usepackage{graphicx}
\begin{document}
\title{Electronic Properties of Clean Au-Graphene Contacts}
%\email[]{dragomir.davidovic@physics.gatech.edu}
%\thanks{}
\author{C. E. Malec and D. Davidovi\'c}

\affiliation{School of Physics, Georgia Institute of Technology,
Atlanta, GA 30332}

%\collaboration{} FG^{as}
\date{\today}

\begin{abstract}

The effects of Au grains on graphene conduction and doping are investigated in this report.
To obtain a clean Au-graphene contact, Au grains are deposited over graphene at elevated temperature and in high vacuum, before any chemical processing. The bulk and the effective contact resistance versus gate voltage demonstrate that Au grains cause p-doping in graphene. The  Fermi level shift is in agreement with first principles calculations, but the equilibrium separation
we find between the graphene and the top-most Au layer is larger than predicted.  Nonequilibrium electron transport displays giant-phonon thresholds observed in graphene tunnel junctions, demonstrating the tunneling nature of the contact, even though there are no dielectrics involved.
\end{abstract}
\maketitle

Graphene's ultrarelativistic electronic energy spectrum near the Fermi level is responsible for many
interesting phenomena in electron transport.~\cite{novoselov,zhang,katselson,young} Low carrier density and high mobility, near the charge neutrality point in graphene, suggest various possibilities for electronic devices. For example, by changing the carrier density with metal contacts, one can make p-n and p-n-p junctions.~\cite{huard,lee,blake,barraza,khomyakov1,xia}
It has been shown from first principle calculations that the electronic structure of graphene is strongly perturbed with
Co, Ni, and Pd contacts, while with Al, Ag, Cu, Au, and Pt contacts, the ultrarelativistic character of the carriers in graphene remains intact.~\cite{giovannetti,khomyakov}  An intuitive explanation for the absence of hybridization between graphene and metal, in the latter case, is that the graphene K,K'-point in the reciprocal space is completely outside the free electron Fermi surface in the metal. The difference between the in-plane wavevectors reduces the coupling between the metal and the graphene wavefunctions. One consequence is a large equilibrium separation between carbon atoms in graphene and metal atoms on the surface of the metal.~\cite{giovannetti,khomyakov} The metal contacts induce charge transfer (doping) in graphene in response to the difference between the work functions. In this letter we investigate clean Au-graphene contacts, and find that graphene is significantly p-doped, more than expected from theory. In addition, we  find that nonequilibrium electron transport through Au-graphene exhibits inelastic thresholds, at the same energy as the inelastic thresholds found in graphene tunnel junctions.~\cite{zhangyb} This demonstrates that the clean Au-graphene contact is similar to a tunnel junction.

We make graphene flakes by mechanical exfoliation from natural graphite.~\cite{novoselov1} As a substrate, we use degenerately doped Si-wafers covered with a 300nm thick thermally grown SiO$_2$ layer. The Si-wafer is used as a back gate. Optical contrast and Raman spectroscopy confirm single layer graphene flakes.~\cite{ferrari}  Prior measurements of the metal-graphene contact resistance
show big variability; the fabrication, temperature, the metal used, and gate voltage seem to have an effect on the contact resistance.~\cite{danneau,russo,nagashio,venugopal,xia}
If lithography is involved between the exfoliation step and the metal deposition step, which appears to have been the case in the prior measurements, then polymer residue can be left in the contact, thereby changing the contact resistance.  To eliminate the residue, the contacts in our samples are made without any lithography. The samples are mounted on a metal deposition stage and pumped to high vacuum (1.0$\times$10$^{-7}$Torr), immediately after the exfoliation before any chemical processing. The samples are heated in high vacuum to 250C for approximately 12 hours. After this bake-out, a Au film is deposited over the sample at $\sim$460$^{\circ}$C at the rate of 1.0 nm/s, by thermal evaporation. The nominal Au film thickness is in the range 15nm-20nm. At this temperature and thickness range, Au forms isolated grains. Fig. 1A displays a sample obtained after depositing 20 nm of Au. There are three regions in the figure, corresponding to three different substrates covered by Au grains: SiO$_2$, a graphene flake, and a graphite flake. The morphology of Au varies among the various substrates, but individual grains are always well separated.  The resistance over the SiO$_2$ is immeasurably large. The largest grains are found over the single layer graphene, where the majority of the surface is covered by grains.
 A zoomed-in region
of the grains is shown in an inset in Fig. 1A. The graphene between the grains makes channels with typical length $L=$35nm and width $w=$160nm, which is between the range previously studied by theory.~\cite{barraza} The grain coverage, as well as the graphene channel dimensions $L$ and $w$, are affected by the amount of deposited Au, as well as the precise temperature, and substrate interactions.

Next, we make Cr/Au electric contacts to those single Au grains that overlap between the SiO$_2$ and the graphene substrate. Figs. 1B and C display the same graphene flake, with a variety of Cr/Au leads in contact with the overlapping grains. As highlighted in Figs. 1B and C, Cr/Au leads approach the overlapping grains from the SiO$_2$ side, without touching the graphene directly. We use standard electron beam lithography, with Poly(methyl methacrylate) (PMMA) resist, Cr/Au deposition by metal evaporation, and lift-off. To align the Cr/Au leads with the grain, various grains are imaged with a scanning electron microscope and registered before lithography. After PMMA deposition and bakeout at 180C, the leads are written with respect to the registered grains as desired, by electron beam lithography. We use grains within the 2$\mu$m$\times$2$\mu$m square near the middle of Fig. 1B as alignment markers.
We confirm that the Cr/Au leads are in good electric contact to the grains,  by measuring the resistance between the leads at low temperature (k$\Omega$s).
Fig. 1D displays a sample with intermediate grain coverage.
The inset displays a control graphene sample, which has no Au grains.
We have studied in detail two samples for each grain coverage, the major results presented here were reproducible among those samples.

Fig. 2 displays four probe bulk resistance  versus gate voltage at 4.2K, in the control sample (Fig. 2A); the sample with intermediate
grain coverage (Fig. 2B); and the sample with high grain coverage (Fig. 2C). The resistance is
measured by lock-in voltage detection, at the excitation current 100nA. As the grain coverage increases, the resistance maximum shifts to higher gate voltage,
reaching $V_{g,max}$=85V in Fig. 2C, indicating p-doping in graphene in accordance with the grain coverage. 
Similarly, the electron-hole asymmetry in the resistance maximum increases with grain coverage. A wider resistance peak in Fig. 2B
with two visible maxima and electron hole asymmetry would be qualitatively consistent with theoretical findings,~\cite{barraza}
discrepancies may be due that in our case electrons can follow a large number of paths and the resistance curve reflects 
some sort of averaging. 

Fig. 3A shows that the bulk resistance in the high grain coverage sample, increases with the perpendicular applied magnetic field. 
In the intermediate grain coverage sample, the resistance increases with magnetic field up to 8T, after which is starts to decrease with the field (not shown). Quantum Hall effect is not yet developed in that sample at 12T.
%The Hall resistance asymmetry about zero magnetic field, in a high grain coverage sample, confirms 
%the charge neutrality point near 90V. 
In the control sample, the four probe longitudinal resistance and the Hall resistance, versus gate voltage and magnetic field, display the half-integer quantum Hall effect as expected in graphene,~\cite{zhang,novoselov} at hole mobility $5,400cm^2/Vs$.

The electron-hole asymmetry in bulk resistance, versus grain coverage, can be explained by p-n junctions, as in Ref.~\cite{huard}. Since the graphene channels in Fig. 1A are short and wide, the channels will be doped because of the proximity to the contact.~\cite{golizadeh,barraza,khomyakov1}
From the calculations in Refs.~\cite{khomyakov1}, we estimate that the charge density near the middle of the channels, at zero gate voltage, is
approximately 50\% of the charge density directly under the contact. At the gate voltage below the resistance maximum, both the channels and the graphene under the contact are p-doped. As the gate voltage increases, the charge neutrality will be reached in the channels first, creating p-n junctions, thereby reducing the slope in bulk resistance versus gate voltage.~\cite{huard}

Next, we measure the effective contact resistance, defined as the ratio of the voltage measured between leads 3 and 4, and the current applied between leads 2 and 1. Fig. 3B displays the gate voltage dependence of the effective contact resistance, versus magnetic field. The resistance maximum is now near 120V, demonstrating that the doping is enhanced compared to the bulk. At 120V, the added electron density in graphene, induced by the gate charge, is $n=C_gV_{g,max}/|e|=9.3\cdot 10^{12}/cm^2$, where $C_g$ is the capacitance to the gate per unit area, measured to be $12.4nF/cm^2$ on a test sample from the same batch of oxidized Si-wafers. This corresponds to the p-doping in graphene with a Fermi level shift $\Delta E_F=\hbar v\sqrt{\pi n}=0.35eV$, where we assume $v=10^6m/s$. First principle calculations of the Fermi level shift under a clean Au-graphene contact, under $\langle 111\rangle$ Au face, predict that $\Delta E_F=0.19eV$ and the equilibrium separation between the carbon atoms in the graphene sheet and the Au atoms of the top-most Au layer 3.3\AA.~\cite{giovannetti,khomyakov}
The calculation leads to $\Delta E_F=0.35 eV$ at the separation of $\approx 4\AA$.~\cite{giovannetti,khomyakov}
%Possible physical origins of the larger separation between Au and the graphene in our sample is discussed in the conclusion.

The observed Fermi level shift is also larger than reported in previous experiments.
The effects of individual Au atom adsorbates on graphene conduction have been investigated
at low temperatures.~\cite{mccreary} Individual Au atoms lead to n-doping in graphene, but as Au-atoms bind into clusters, the Fermi level shifts back to neutrality.~\cite{mccreary} The measurements of $\Delta E_F$ in large Ti/Au-graphene contacts, obtain $\Delta E_F=0.25eV$ by photocurrent microscopy,~\cite{lee} but those contacts involved electron-beam lithography over graphene, before the metal deposition. Photoemission spectroscopy of SiC-graphene with intercalation of Au monolayers displayed smaller p doping, $\Delta E_F=0.19eV$.~\cite{varykhalov}

At zero gate voltage, the effective contact resistance is $915\Omega$.  The contact area between the grain and graphene, estimated from the sample image, is $\approx 0.016\mu m^2$, so the effective resistivity of the contact would be $\rho= 14.6\cdot 10^{-8}\Omega cm^2$. Alternatively, the diameter of the grain is approximately 140nm, and so the effective resistance per unit length is only $128\Omega\mu m$, comparable to the current record.~\cite{xia} At -100V on the back gate, the contact specific contact resistance drops to $95\Omega\mu m$.  The effective contact resistance measured in other similarly sized grains agrees with the above.
The effective contact resistance is equal to the contact resistance only if the spread resistance from graphene under the contact and from graphene surrounding the contact is negligibly small compared to the contact resistance.
If the spread resistance is significant, then the effective contact resistance will be larger than the contact resistance.
Thus, the estimate presents an upper bound of the contact resistance. We expect that effective contact resistance is not far above the contact resistance.
The contribution to the effective contact resistance from the graphene channels cannot be strong, because of the very weak magnetic field dependence of the effective contact resistance compared to the bulk (Figs. 3A and B). The graphene channels are less doped than the graphene directly under the contact. Since the maximum in the effective contact resistance is 30V above the bulk resistance maximum,
the channels have a reduced contribution to the effective contact resistance. 

Next, we discuss nonequilibrium electron transport. Figs. 4 A and B display bulk differential resistance versus bias voltage at 4.2K, in samples with intermediate and  high grain coverage, respectively. In this measurement, a 4 probe resistance measurement of the graphene/Au grain system is measured and graphed versus the DC bias voltage applied to the current source leads.  The dominant feature in the figure is a resistance maximum near zero bias voltage, or a zero-bias anomaly (ZBA). The ZBA is common in mesoscopic electron transport, and generally arises from the enhancement of electron-electron interactions in
samples with restricted dimensions and weak disorder.~\cite{altshuler,devoret,nagaev} Similar ZBA is confirmed in the control sample, although the ZBA versus magnetic field, in the control sample exhibits oscillations between resistance maxima and the resistance minima,
due to the quantum Hall effect. The discussion of the relation between the ZBA and the quantum Hall effect is outside the scope of this report.

The ZBA in Au-covered graphene samples exhibits additional peaks. The first peak is observed near $\pm$70mV, as indicated by arrows in Fig. 3. All the peak locations are symmetric with respect to the sign of the bias voltage. The absence of peaks below 70mV suggest that the sample resistance is affected by some inelastic scattering process requiring an energy difference of at least  70meV.
Inelastic conduction thresholds near 70meV have been observed in scanning tunneling spectroscopy in graphene~\cite{zhangyb} and in graphene tunnel junctions.~\cite{malec} They have been attributed to the 67 meV out-of-plane acoustic
graphene phonon modes located near the K/K' points in reciprocal
space.~\cite{mohr} Electrons with energy less than this phonon energy tunnel elastically between graphene and the metal.
Due to the conservation of momentum, the
effective barrier height of the tunneling junction
is enhanced by $\hbar^2K^2/2m\approx 11eV$, so the probability of the elastic tunneling is reduced.~\cite{tersoff,tersoff1,vitali}
In inelastic tunneling, an electron at energy 67meV above the Fermi level can tunnel through the barrier with zero in plane momentum, through the emission of a K point out-of-plane phonon.~\cite{zhangyb} The barrier height for the inelastic tunneling is reduced by $11eV$
compared to that for the elastic tunneling, enhancing the probability of inelastic tunneling.

In absence of electron-phonon relaxation in our
samples, the electron distribution will have an effective temperature $eV/k_B$, where $V$ is the bias voltage. At bias voltage above 67mV, an electron can make a tunneling transition between graphene and a Au grain with zero in plane momentum, through the emission of a K point out-of-plane phonon.  The probability of electron tunneling across the Au-graphene interface is enhanced when the available electron energy exceeds 67meV.  Thus, the inelastic threshold near 70mV suggest that the clean Au-graphene contact is a tunnel junction. The barrier in this junction would be purely kinetic, that is, there is no requirement for a dielectric inside the barrier. The existence of this kinetic barrier suggests a possibility for a graphene transistor
without any dielectric separating the gate metal and the graphene.

In conclusion, Au grains in clean contact with graphene lead to significant p-doping, with the Fermi level shift $\Delta E_F=0.35eV$, consistent with first principle calculations if the separation between the the graphene layer and the Au layer closest to the graphene is greater by $\approx 1\AA$ with respect to equilibrium on the $\langle 111\rangle$ face.. The substrate interaction can affect the separation between carbon and Au atoms. Nonuniform conformation of the substrate may play a big role. Van der Waals contributions from the SiO$_2$ substrate, which are not included in the first principle calculations, may also induce large graphene-metal separations. Nonequilibrium electron transport in Au-covered graphene exhibits inelastic thresholds at the giant phonon energy in graphene, confirming the tunneling nature of the
Au-graphene contact. This tunneling occurs even though there are no dielectrics in the contact. This work was supported by the department of energy (DE-FG02-06ER46281). We thank S. Barraza-Lopez, M. Kindermann, and M. Y. Chou for valuable discussions.

\bibliography{graphene}
\newpage

\begin{figure}%[p]
\begin{center}
\includegraphics[width=0.75\columnwidth]{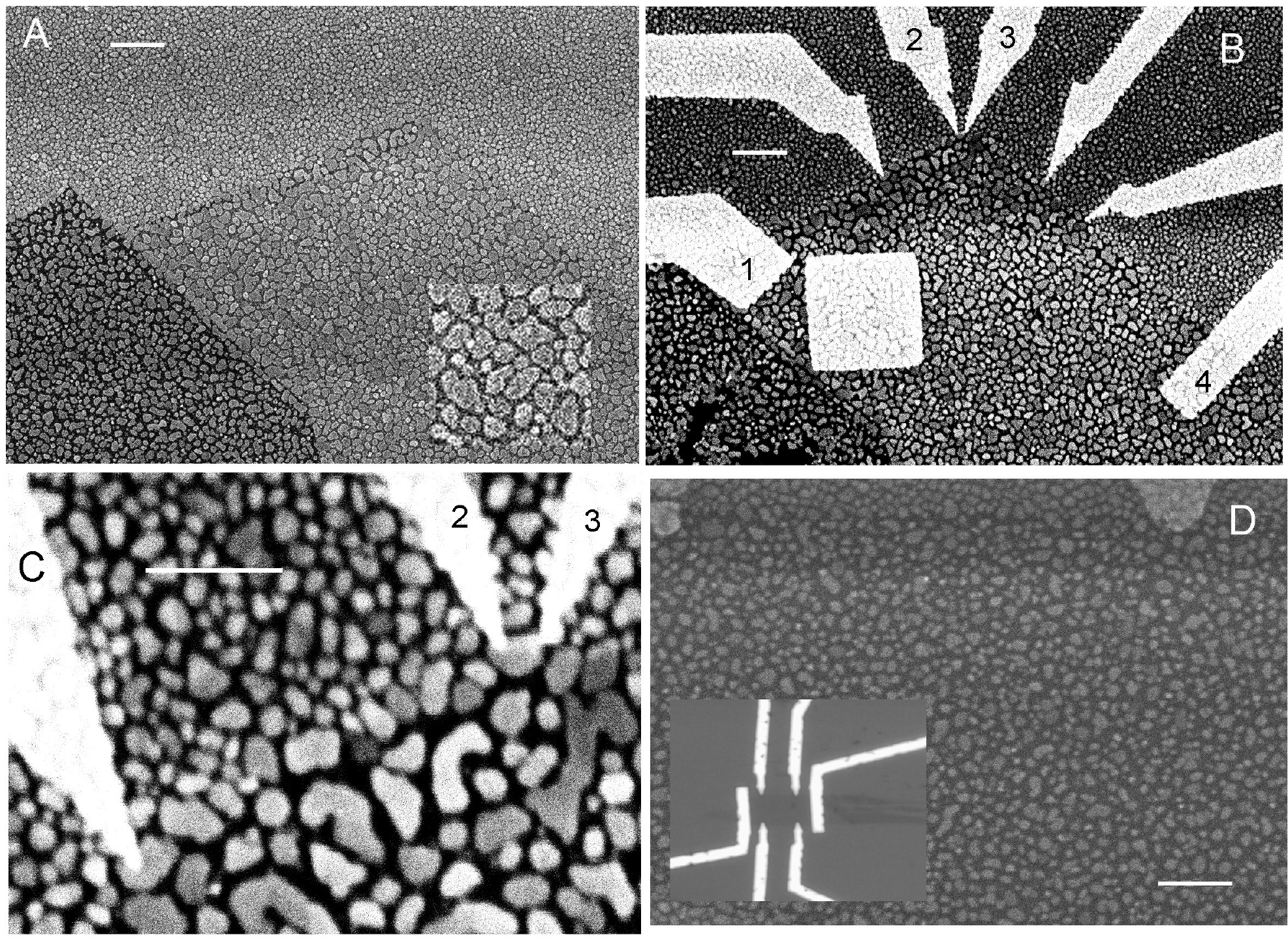}
 \caption{A: Scanning electron micrograph of SiO$_2$, graphene, and graphite, covered with Au grains, in a high grain coverage sample. Inset:  A square micron of grains over graphene. B: The same as in A, after Cr/Au lead deposition. C: Leads 2 and 3 from B, zoomed-in. D: Graphene with intermediate grain coverage. Inset: Optical image of a control sample. The width of the graphene flake is 4$\mu$m. The white bars in A-B and C-D correspond to 1$\mu$m and 0.5$\mu$m, respectively.}
\end{center}
\end{figure}

\begin{figure}%[p]
\includegraphics[width=0.75\columnwidth]{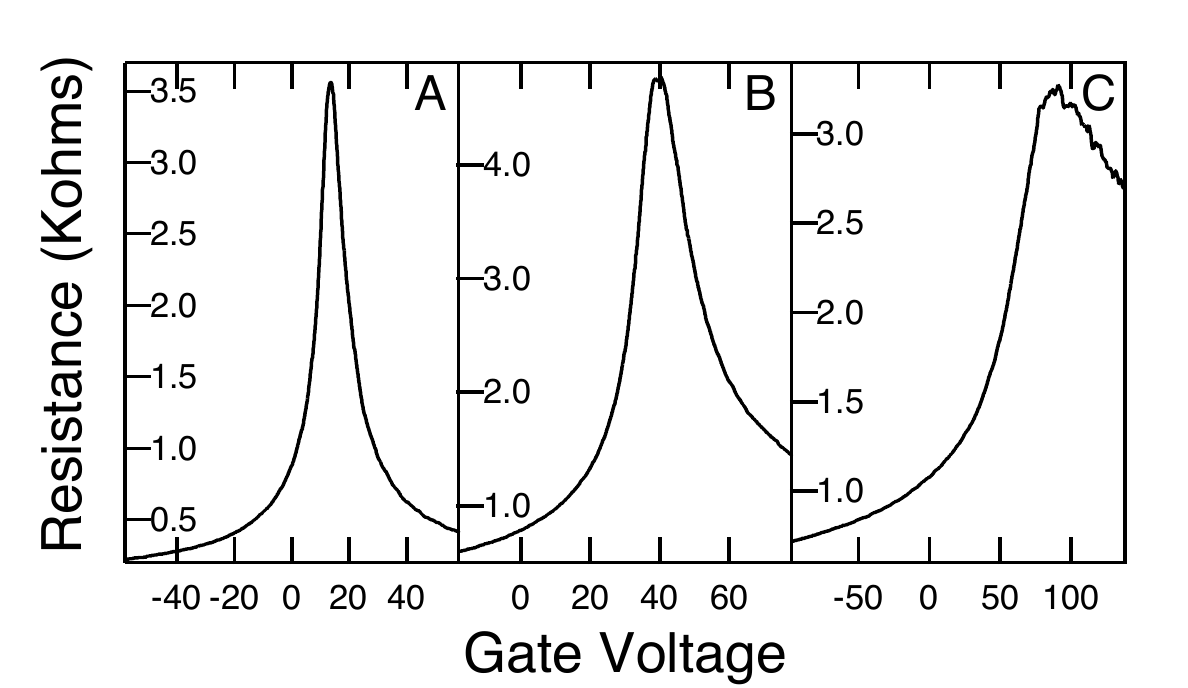}
\caption{A, B, C: Bulk four-probe resistance versus gate voltage, for a control sample, a sample with intermediate grain coverage, and a sample with high grain coverage, respectively. T=4.2K.}
\end{figure}

\begin{figure}%[p]
\includegraphics[width=0.75\columnwidth]{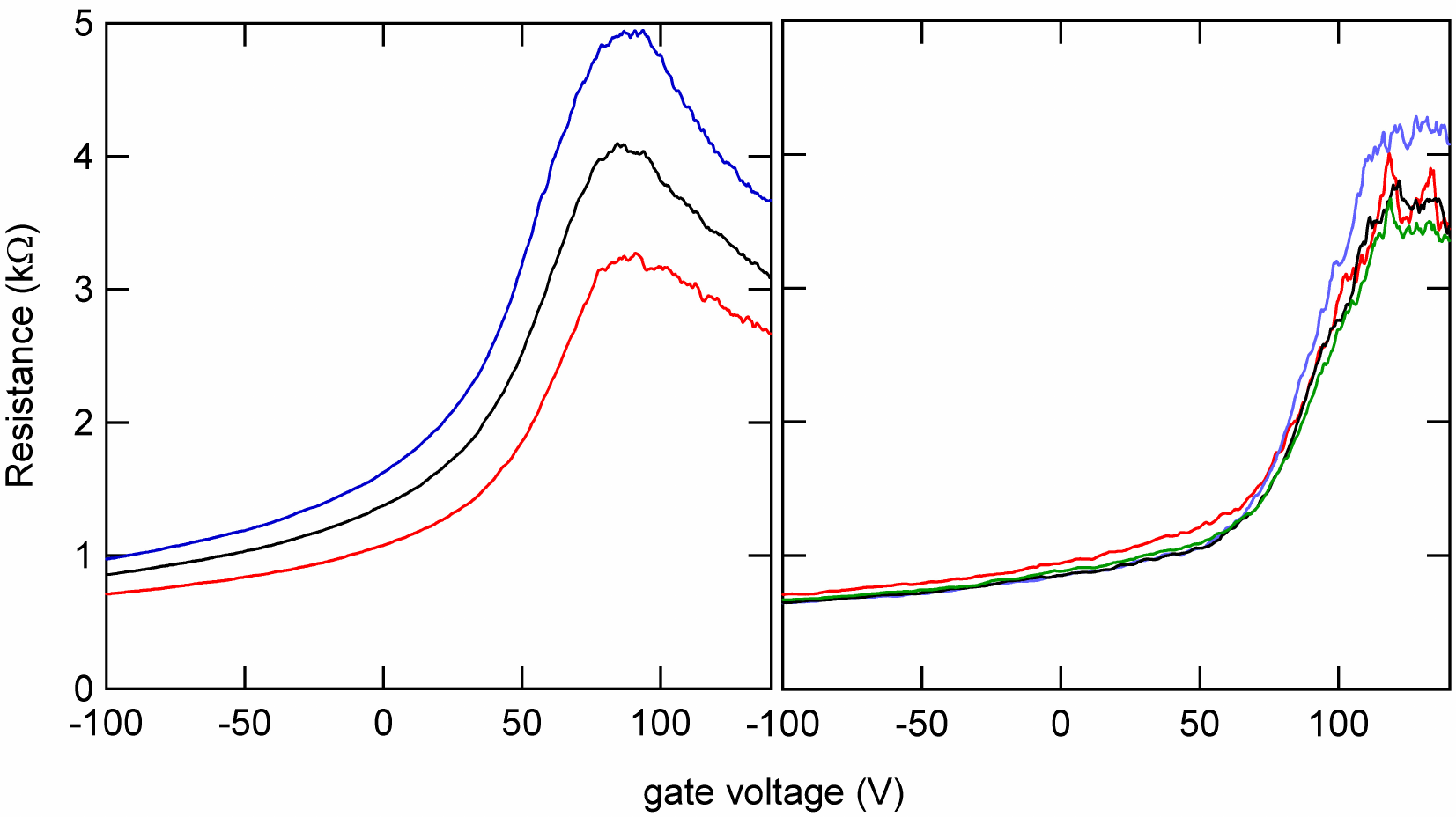}
 \caption{A. Four probe sample resistance versus gate voltage. Red, black, and blue measured at 0T, 8T, and 12T, respectively.  B. The effective contact resistance between the grain and graphene, versus gate voltage. Red, green, black, and blue measured at 0T, 4T, 8T, and 12T, respectively. T=4.2K.}
\end{figure}

\begin{figure}%[p]
\includegraphics[width=0.75\columnwidth]{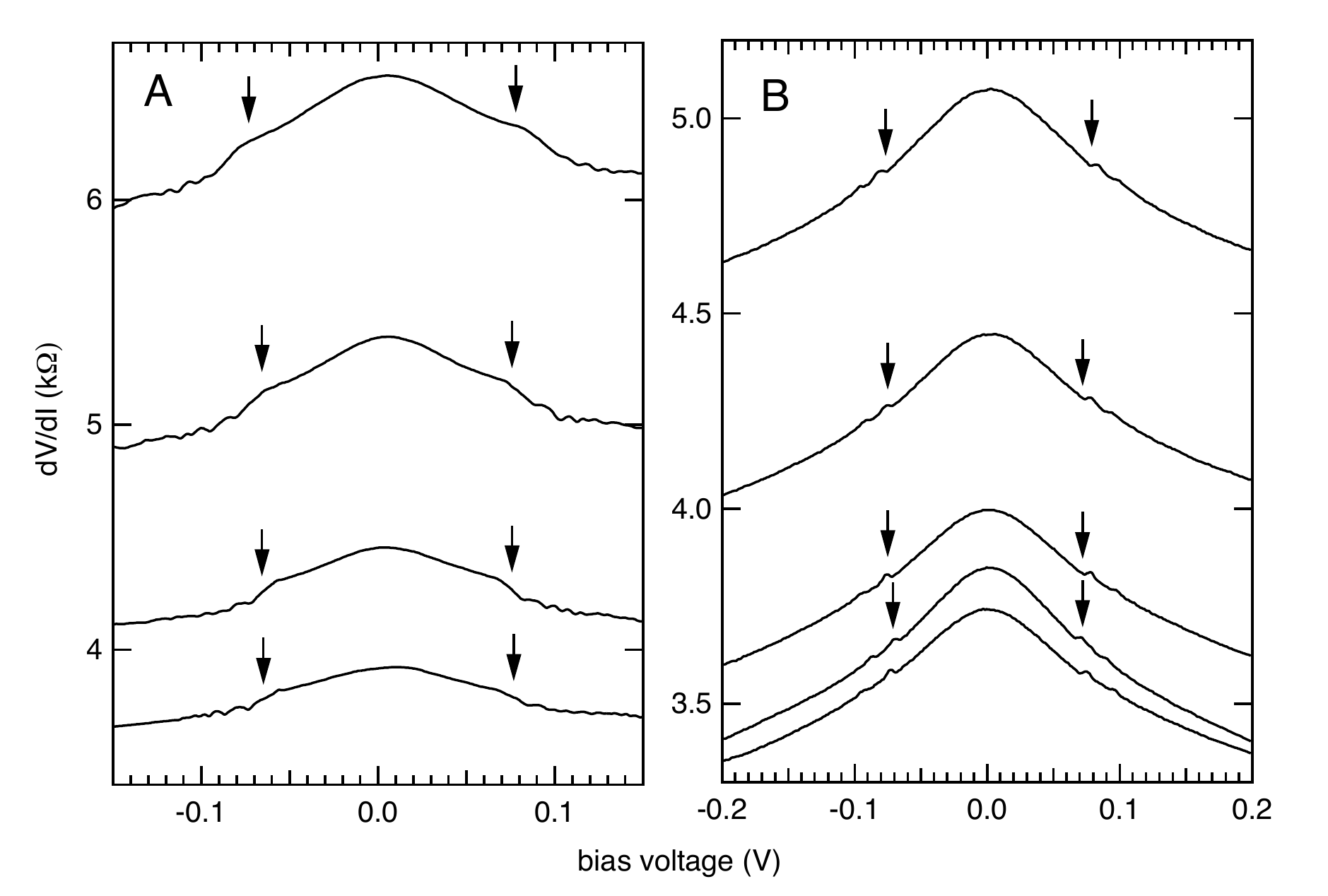}
 \caption{A and B: Differential resistance versus bias voltage, for intermediate and dense grain coverage, respectively. Gate voltage 50V and 100V, in A and B, respectively. Both are approximately 15V above the Dirac point.  In A, the applied magnetic field 0, 2T, 4T, and 6T, bottom to top. In B, the applied magnetic field 0, 2T, 4T, 6T, and 8T, bottom to top. T=4.2K.}
\end{figure}
\end{document}